\documentclass{ws-ijmpd}

\begin{document}

\markboth{Eduardo S. Pereira; Oswaldo D. Miranda}
{Massive Black Hole Binary Systems in  Hierarchical Scenario of Structure Formation}

\catchline{}{}{}{}{}
%
%%%%%%%%%%%%%%%%%%%%%%%%%%%%%%%%%%%%%%%%%%%%%%%%%%%%%%%%%%%%%%%%%%%%

\title{Massive Black Hole Binary Systems in  Hierarchical Scenario of Structure Formation}

\author{EDUARDO DOS SANTOS PEREIRA}

\address{INSTITUTO NACIONAL DE PESQUISAS ESPACIAIS - INPE,
Av. dos Astronautas,1.758 - Jd. Granja, Sao Jose dos Campos, SP, 12227-010,
Brazil, duducosmos@das.inpe.br}

\author{OSWALDO D. MIRANDA}

\address{INSTITUTO NACIONAL DE PESQUISAS ESPACIAIS - INPE,
Av. dos Astronautas,1.758 - Jd. Granja, Sao Jose dos Campos, SP, 12227-010,
Brazil, oswaldo@das.inpe.br}

\maketitle

\begin{history}
\received{Day Month Year}
\revised{Day Month Year}
\comby{Managing Editor}
\end{history}

\begin{abstract}
The hierarchical scenario of structure formation describes how objects like galaxies and
galaxy clusters are formed by mergers of small objects. In this scenario, mergers of
galaxies can lead to the formation of massive black hole (MBH) binary systems. On the
other hand, the merger of two MBH could produce a gravitational wave signal detectable,
in principle, by the Laser Interferometer Space Antenna (LISA). In the present work,
we use the Press?Schechter formalism, and its extension, to describe the merger rate of
haloes which contain massive black holes. Here, we do not study the gravitational wave
emission of these systems. However, we present an initial study to determine the number
of systems formed via mergers that could permit, in a future extension of this work, the
calculation of the signature in gravitational waves of these systems.
\end{abstract}
 

\section{Introduction}	
Recently, observational evidence for the existence of massive black holes in galaxies
has been reported in the literature. On the other hand, the hierarchical scenario
of structure formation describes how objects like galaxies and galaxy clusters are
formed in the early universe. In this way, we can suppose that mergers of galaxies
can lead to the formation of the massive black holes (MBH) observed in galaxies and
their binary systems. Thus, the main goal of this work is to describe a method to
determine the evolution of binary systems of massive black holes in the hierarchical
scenario. To do that, we present in Sec. 2 a short review on the Press?Schechter
formalism. In Sec. 3 we describe how to obtain the relation between the central
black hole and the mass of the host dark halo. We also present the way to calculate
the number of massive binary systems in Sec. 3. In Sec. 4 we present the main
results, and finally in Sec. 5 we present our conclusions. Our models are obtained
using the following set of cosmological parameters: $\Omega_{m} = 0.24$, $\Omega_{b}=0.04$, $\Omega_{\Lambda}=0.76$ and $h=0.73$.

\section{Hierarchical scenario of structure formation: Press-Schechter formalism and its extension}
The hierarchical scenario of structure formation: the core of Press-Schechter \cite{ps} (P-S). formalism is that a dark matter halo leaves the linear regime when the mean density within a given volume is larger than a threshold level  $\delta_{c}$. 
In particular, Lacey and Cole \cite{lc} proposed an extension of the P-S formalism based on the Brownian
random wake of  Bond et al.\cite{bce} . The goal of this extension was to take into account the probability that a dark matter halo (henceforth halo), with mass  $M_{1}$, has to merger with another halo with  mass $M_{2}$, for any redshift z, in order to form a new halo with mass $M_{f}= M_{1}+M_{2}$. Fakhouri and Ma\cite{fama} showed that, using the P-S formalism and its extension, the merger rate of haloes is given by:
\begin{equation}\label{tfu}
\frac{B(M_{1},M_{f},z)}{f(M_{f}, z ;P-S) } = \sqrt{\frac{2}{\pi}} \frac{1}{\sigma^{2}(M_{1})} \left| \frac{d \sigma(M_{1})}{dln(M_{1})}\right| \left| \frac{d \delta_{c}}{dz}\right| \left[1 -\frac{\sigma^{2}(M_{f})}{\sigma^{2}(M_{1})} \right]^{-3/2}
\end{equation}

\noindent where $f(M,z;P-S)$ is the P-S mass function of dark haloes, $\sigma(M)$ is the variance and $B(M_{1},M_{f},z)$ is the merger of dark halos.

\section{Binary systems of massive black holes }
Wythe \& Loeb\cite{wl}  proposed a model for the relation between central black hole and the mass of host dark halo.
The authors consider that the central black hole (CBH) stop  growing when the accretion  reaches
the Eddington luminosity. In particular, they  consider that the circular velocity is equal the virial velocity, in this case, the mass of dark halo $M_{h}$, as a function of the CBH is (see Refs \refcite{wl}, \refcite{eric}):
\begin{equation}\label{mhalo0}
M_{h}(M_{BH}) = \varepsilon_{0}^{-3/5} \left( \frac{\Omega^{0}_{m}}{\Omega_{m}(z)} \frac{\Delta_{c}}{18 \pi^{2}} \right)(1+z)^{-3/2} \left( \frac{M_{BH}}{10^{12}\rm{M}_{\odot}} \right) 10^{12} \rm{M}_{\odot},
\end{equation}

\noindent where $\Omega^{0}_{m}$ is the dark matter parameter at present time, $\varepsilon_{0} = 10^{-5,7}$ and $\Delta_{c}$ 
is the linear overdensity by virialization of a spherical pertubation ``\textit{top-hat}''-like,  that for $\Lambda$CDM 
is\cite{lmbl}:
\begin{equation}
\Delta_{c} = 18 \pi^{2} + 82[\Omega_{m}(z)-1] -39[\Omega_{m}(z) - 1]^{2}.
\end{equation}

If we consider that the fraction $\epsilon_{1}$ of dark haloes, at $z< 10$, having a central MBH,then we can obtain, using  equations (\ref{mhalo0}) and (\ref{tfu}), the following equation to the formation rate of massive binary systems:
\begin{equation}\label{bfr}
R(M_{BH,1},M_{BH,2},z) = \epsilon_{1} \epsilon_{2} f(a) B(M_{h}(M_{BH,1}),M_{h}(M_{BH,2}),z),
\end{equation}

\noindent where $M_{bh,i}$ $(i=1,2)$ is the mass of CBH, $a$ is the separation of binary system at $z$ and $f(a)$ is the separation distribution 
function.

We considered in Eq. (\ref{bfr}) that the central MBH quickly forms binary systems
and which the fraction $\epsilon_{2}$ of these systems achieve the gravitational wave regime.
On the other hand, for the separation distribution function we have:
\cite{hs}:
\begin{equation}
f(a)da = \frac{3}{2} \left[ \left( \frac{a}{\overline{x}}\right)^{3/4}-\left(\frac{a}{\overline{x}} \right)^{3/2} \right]\frac{da}{a}.
\end{equation}
In the above equation we consider that the MBHs are formed in galaxy clusters.
Thus, $\overline{x}$ represents the maximum separation of the components, and $\overline{x} $ is the typical dimension of galaxies clusters. 

The  density  number ($n_{BH}$) of black hole binary systems obeys the conservation equation:
\begin{equation}\label{dnsy}
\frac{\partial n_{BH}}{\partial z}\left| \frac{dz}{dt}\right| + \frac{\partial(n_{BH}(da/dt))}{\partial a} = R(M_{bh,1},M_{bh,2},z).
\end{equation}

In a gravitational wave regime, the variation of separation with time is:
\begin{equation}
\frac{da}{dt} = - \frac{64}{5} \frac{G^{3}}{c^{5}}\frac{(M_{BH,1}+M_{BH,2})}{a^{3}}M_{BH,1}M_{BH,2}.
\end{equation}

Finally, the number of systems, as a function of $z$ and observed frequency, $\nu_{obs}$, is:
\begin{equation}
N_{sys} = - n_{BH}\frac{dV}{dz}\frac{da}{d\nu_{obs}}.
\end{equation}

\noindent with:
\begin{equation}
\frac{da}{d\nu_{obs}} = -\frac{3}{2\pi}\left[ G(M_{BH,1}+M_{BH,2})\right]^{1/2}a^{-5/2}(1+z)
\end{equation}

\noindent where $dV/dz$ is the comovel volume.

\section{Numerical results}

\begin{figure}
%\center
      \includegraphics[height=45mm,width=130mm]{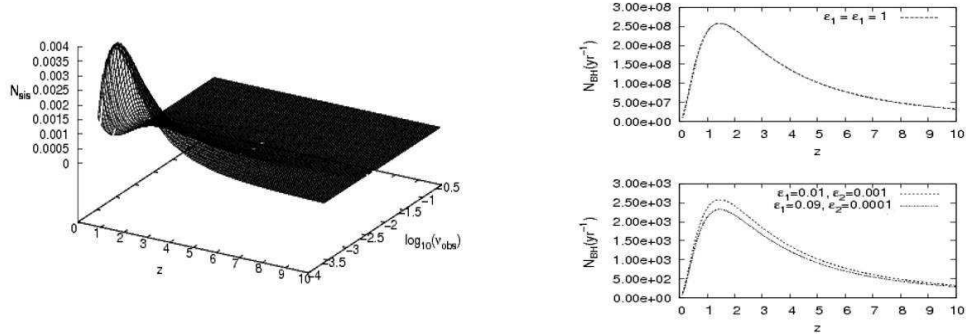}
\vspace*{8pt}       
\caption{The left hand side is showed the number of systems as a function of observed frequency and z with $\epsilon_{1}=\epsilon_{2}=1$ and black holes with mass $M_{BH,1}=10^{5}$ and $M_{BH,2}=0.1M_{BH,1}$. In the 
right hand right, is presented the total number of binary sistems of massive black holes into the mass range $10^{4}\rm{M}_{\odot} \leq M_{BH,1} \leq 10^{7}\rm{M}_{\odot}$ and $0.1M_{BH,1}\leq M_{BH,2} \leq M_{BH,1}$.
On the top it was assumed $\epsilon_{1}=\epsilon_{2}=1$ and on the botton it was considered $\epsilon_{1}=0.01$, $\epsilon_{2}=0.001$, $\epsilon_{1}=0.09$ and $\epsilon_{2}=0.0001$.
}\label{fig:1}
            \end{figure}

In this work, we use the Lax-Wendroff schema\cite{ptv} to obtain the numerical solution of equation (\ref{dnsy}). Figure \ref{fig:1}, on 
the left hand side, shows the number of systems, by frequency by redshift, of binary system formed by black holes of mass
 $M_{BH,1} = 10^{5} \rm{M}_{\odot}$ and $M_{BH,2} =0.10M_{BH,1}$ and, on the right hand side, the total number of systems into the mass range 
$10^{4}\rm{M}_{\odot} \leq M_{BH,1} \leq 10^{7}\rm{M}_{\odot}$ and $0.1M_{BH,1}\leq M_{BH,2} \leq M_{BH,1}$ for different values of
 $\epsilon_{1}$ and $\epsilon_{2}$.In both cases, we assumed that the separation range is $3(r_{sh,1}+r_{sh,2})\leq a\leq 100(r_{sh,1}+r_{sh,2})$, 
with $r_{sh,i}$ is the Schwardschild ratio of black hole $i$ and we assumed $\overline{x}=1.5~\rm{Mpc}$. Note that the function $N_{BH}$ peaks at $z \approx 2$.

\section{Conclusion}

We presented a different method to calculate the number of binary systems of mas-
sive black holes using the Press-Schechter formalism. It is important to emphasize
that the total number of systems obtained here takes into account all systems. This
is a different result when compared, for example, for the work of
  Wythe \& Loeb \cite{wl}  who obtained their results only for systems within the LISA range. The numeri-
cal method used here is stable and it produces a smooth function for the density
of binary systems. The same method was used by
 Banerjee \& Ghosh\cite{bach} or the
calculus of the formation of binary systems in globular clusters of stars.

\section*{Acknowledgments}
E. S. Pereira would like to thank the Brazilian Agencies CAPES and FAPESP for supporting, also Dr. Odylio. D. Aguiar.
O. D. Miranda would like to thank the Brazilian Agency CNPq for partial support (grant 305456/2006-7).

%\appendix

%\section{Appendices}

%\section{References}

%\begin{thebibliography}{000} %for 3 digits
%\begin{thebibliography}{00}  %for 2 digits

\end{document}